\documentclass[journal]{IEEEtran}

\usepackage{graphicx}
\usepackage{dcolumn}
\usepackage{bm}
\usepackage{caption}
\usepackage{comment}
\usepackage{subcaption}
\usepackage{xcolor}
\usepackage{physics}
\usepackage{ulem}
\usepackage{placeins}
\usepackage{amsmath}
\usepackage{cite}
\usepackage{verbatim}

 \usepackage{algpseudocode}

\usepackage{algorithm}

\begin{document}

\title{An Envelope Tracking Approach for Particle in Cell Simulations}
\author{O. H. Ramachandran,~\IEEEmembership{Student Member,~IEEE,}
        Z. D. Crawford,~\IEEEmembership{Student Member,~IEEE,}
        S. O'Connor,~\IEEEmembership{Student Member,~IEEE,}
        J. Luginsland,~\IEEEmembership{Fellow,~IEEE}
        B. Shanker,~\IEEEmembership{Fellow,~IEEE}
\thanks{ O. H. Ramachandran, Z. D. Crawford, S. O'Connor and B. Shanker are with the Department
of Electrical and Computer Engineering, Michigan State University, East Lansing,
MI, 48824.\protect\\
 J. Luginsland was with MSU's Electrical and Computer Engineering Department.  He is current with AFRL/Air Force Office of Scientific Research, Arlington, VA 22201.
E-mail: ramach21@egr.msu.edu}
}
\maketitle

\begin{abstract}
The state of the art in electromagnetic Finite Element Particle-in-Cell (EM-FEMPIC) has advanced significantly in the last few years; these have included understanding function spaces that must be used to represent sources and fields consistently, and how currents should be evolved in space and time. In concert, these achieve satisfaction of Gauss' laws. All of these, were restricted to conditionally stable explicit time stepping. More recently, there has been advances to the state of art: It is now possible to use a implicit EM-FEMPIC method while satisfying Gauss' law to machine precision. This enables choosing time step sizes dictated by physics as opposed to geometry. In this paper, we take this a step further. For devices characterized by a narrowband high frequency response, choosing a time-step size based on the highest frequency of interest is  considerably expensive. In this paper, we use methods derived from envelope tracking to construct an EM-FEMPIC method that analytically provides for the high-frequency oscillations of the system, allowing for analysis at considerable coarser time-step sizes even in the presence of non-linear effects from active media such as plasmas. 
Consequentially, we demonstrate how the pointwise metric used for measuring satisfaction of Gauss' Laws breaks down when prescribing analytical fast fields and provide a thorough analysis of how charge conservation can be measured. 
Through a number of examples, we demonstrate that the proposed approach retains the accuracy the regular scheme while requiring far fewer time steps. 
\end{abstract}

\section{Introduction \label{sec:introduction}}
The simulation of moving charges in an electromagnetic field distribution is of great interest in a number of engineering applications. These include particle accelerators, plasma processing applications such as sterilization of medical implements and etching of high precision integrated chips \cite{marchand2011ptetra,lemke1999three,fourkal2002particle}, among many others. The foremost application in performing these simulations is the Particle-in-Cell method (PIC) that relies on self-consistently solving Maxwell's equations for the fields and Newton equations for particles of the charged species.
While traditional implementations of PIC have used finite difference time domain (FDTD) to evolve the fields \cite{verboncoeur2005particle}, recent advances have made viable the use of Finite Element based methods \cite{pinto2014charge,moon2014exact,squire2012geometric,moon2015exact,o2021set}, allowing for better resolution of curvilinear grids and the use of more complex function spaces. In addition, work has been done on constructing a structure-preserving FEMPIC scheme \cite{kraus2017gempic,jianyuan2018structure} based on Whitney forms defined by B-splines \cite{buffa2010isogeometric}.
Furthermore, it is possible to sidestep the Courant–Friedrichs–Lewy (CFL) constraint on the timestep size by using unconditionally stable implicit time marching schemes while conserving charge to machine precision \cite{Oconnor_time_integration}.
And finally, techniques have been very recently developed that allow for explicit satisfaction of the Coulomb Gauge, thus alleviating the problem of spurious null-space solutions that exist in most implicit implementations \cite{oconnor2021quasihelmholtz}.

All-in-all, advances in EM-FEMPIC have made it possible to (1) simulate plasma phenomena while capturing the underlying geometry to very good precision and (2) analyse the system in time at the frequency of interest for capturing the desired physics as opposed to the satisfaction of mesh-based stability criteria.
Despite the massive progress that has been accomplished in recent years, there are still classes of devices for which the current state-of-the-art is inefficient to capture for large scale problems.
One such example is the analysis of devices that have a narrowband, high-frequency response. Similarly, devices with features smaller than the electromagnetic wavelength that are nonetheless important for the electrostatic physics are computationally untenable to solve with contemporary techniques.
Despite having a relatively small window of significant frequency content, to simulate such a device, one would need to discretize the simulation in time at a rate determined by the maximal frequency of interest. 

Efficient ways to simulate these devices exist when one considers a purely electromagnetic analysis (i.e without particles).
Termed envelope tracking, these methods solve an altered form of Maxwell's equations wherein the high frequency offset present in the fields is analytically prescribed. 
Performing this operation, effectively a Hilbert Transform of Faraday's and Ampere's laws, yields a new set of equations that still vary in time, but can be discretized at the bandwidth of the signal, which is significantly lower than the maxmial frequency in the unshifted system.

Extending these results to a system with moving macroparticles is challenging for two primary reasons: (1) While the fields themselves oscillate within a narrowband high frequency window, the particle positions and velocities generally have significant DC components that force analysis of Newton's equations at the highest frequency of interest. (2) While recent advances make it possible to conserve charge within an implicit time marching scheme \cite{Oconnor_time_integration}, these techniques break down when used with an analytically prescribed plane wave component (We discuss this in greater detail in Section \ref{sec:ccons_metric}).

In the remainder of this paper, we construct an EM-FEMPIC method wherein the fields are solved for using an envelope tracking method, henceforth referred to as ET-FEMPIC. We address both of the challenges listed above by (1) constructing an integration scheme for the particle system that allows Newton's laws to be solved at the highest frequency of interest while being mapped self consistently with the Maxwell solver and (2) by constructing a Quasi-Helmholtz framework to explicitly satisfy the Coulomb gauge. 
We show analytically and through numerical experiments that this framework conserves charge to machine precision while producing results that match a traditional EM-FEMPIC implementation. 

\section{Problem Statement \label{sec:probstatement}}

Consider a region $\Omega \in \mathbf{R}^{3}$ bounded by a surface $\partial \Omega$ containing a single charged species. This region is subjected to an external field due to which the charged species accelerate, and in turn produce spatially and temporally varying electric and magnetic fields denoted by $\mathbf{E}(\mathbf{r},t)$ and $\mathbf{B}(\mathbf{r},t)$, respectively, with $\mathbf{r}\in\Omega$ and $t\in[0,\infty)$. The dynamics of the particles in phase space can be represented by a distribution function (PSDF) $f(t,\mathbf{r},\mathbf{v})$ that follows the Vlasov equation:
\begin{equation}\label{eq:distFn}
\begin{split}
    \partial_{t} f(t,\mathbf{r},\mathbf{v}) +\mathbf{v} \cdot \nabla f(t,\mathbf{r},\mathbf{v}) +& \\
    \frac{q}{m}\left[\mathbf{E}(\mathbf{r},t)+\mathbf{v}\times\mathbf{B}(\mathbf{r},t)\right]\cdot \nabla_{v}f(t,\mathbf{r},\mathbf{v}) =& 0.
\end{split}
\end{equation}
In what follows, we assume that the background media in $\Omega$ is free space. As a result, we denote the permittivity and permeability of free space by $\epsilon_{0}$ and $\mu_{0}$, respectively, and the speed of light by   $c$. In what follows, we will assume that either the  external fields  impressed on $\Omega$ or the source exciting fields is narrowband in that the center frequency of the excitation $f_0 \gg f_{bw}$, where $f_{bw}$ is the bandwidth. Likewise, we will assume that the system is quiescent for $t\le 0$.

\section{Formulation \label{sec:formulation}}

We follow the usual path of not solving \eqref{eq:distFn} directly, instead representing the moments of the PSFD using a  charge and current density as, $\rho(t,\vb{r}) = q \int_\Omega f(t,\vb{r},\vb{v}) d\vb{v}$ and $\vb{J}(t,\vb{r}) = q\int_\Omega \vb{v}f(t,\vb{r},\vb{v})d\vb{v}$. We then use a particle approximation of these moments, and evolve their location and velocity together with Maxwell's equations. As a result, assuming a  shape functions $S(\vb{r})$, one obtains
\begin{subequations}
\begin{equation}
    \rho(t,\mathbf{r}) = q\sum_{p=1}^{N_p} S(\mathbf{r}-\mathbf{r}_p(t)),
\end{equation}
\begin{equation}
    \mathbf{J}(t,\mathbf{r}) = q\sum_{p=1}^{N_p} \mathbf{v}_p(t)S(\mathbf{r}-\mathbf{r}_p(t)).
\end{equation}
\end{subequations}
where $N_p$ denotes the number of macroparticles. Furthermore, $\mathbf{r}_{p}(t)$ and $\mathbf{v}_{p}(t)$ refer to the positions and velocities as a function of time of the $p$th macroparticle. With no loss of generality, we will assume that there exist a source $\vb{J}_i(\vb{r}_{s},t) = \tilde{\vb{J}}_i(\vb{r}_{s},t) e^{j \omega_0 t}$ at points $\vb{r}_{s}$ that excites the system. Given this temporal dependence, we posit that the field and fluxes have a similar behavior, i.e.,  $\vb{E}(\vb{r},t) = \tilde{\mathbf{E}}(\mathbf{r},t)e^{j \omega_0 t} $ and $\vb{B}(\vb{r},t) = \tilde{\mathbf{B}}(\mathbf{r},t)e^{j \omega_0 t} $. In these expressions, the quantities with a \emph{tilde} are slowly varying with respect to time. Using these expressions in Maxwell's equations, we can write
\begin{subequations}
\label{eq:maxwell}
\begin{equation}
    \nabla\times\tilde{\mathbf{E}}(\mathbf{r},t) = -j\omega_{0}\tilde{\mathbf{B}}(\mathbf{r},t) - \partial_t \tilde{\mathbf{B}}(\mathbf{r},t),
\end{equation}
\begin{equation}\label{eq:ampere}
\begin{split}
    \nabla\times\mu_0^{-1}\tilde{\mathbf{B}}(\mathbf{r},t) & = \tilde{\vb{J}}_i(\mathbf{r}_{s},t) +  \mathbf{J}(\mathbf{r},t)e^{-j\omega_{0}t} \\
    & + j\omega_{0}\epsilon_0\tilde{\mathbf{E}}(\mathbf{r},t)+\epsilon_0 \partial_t \tilde{\mathbf{E}}(\mathbf{r},t),
\end{split}
\end{equation}
Following \cite{Oconnor_time_integration}, we replace $\mathbf{J}(\mathbf{r},t)$ with its time integral
\begin{equation}\label{eq:g_def}
\begin{split}
\mathbf{G}(\mathbf{r},t) = \intop_{0}^{t} \mathbf{J}(\mathbf{r},\tau) d\tau &= q\sum_{p=1}^{N_{p}}\intop_{0}^{t} \mathbf{v}_{p}(\tau) \delta(\mathbf{r}-\mathbf{r}_{p}(\tau)) d\tau \\
&= q \sum_{p=1}^{N_{p}} \intop_{\mathbf{r}_{p}(0)}^{\mathbf{r}_{p}(t)} d\tilde{\mathbf{r}} \delta(\mathbf{r}-\tilde{\mathbf{r}})
\end{split}
\end{equation} in \eqref{eq:ampere}, to obtain
\begin{equation}
\begin{split}
    \nabla\times\mu_0^{-1}\tilde{\mathbf{B}}(\mathbf{r},t) & = \tilde{\vb{J}}_i(\mathbf{r}_{s},t) +  e^{-j\omega_{0}t}\partial_{t}\mathbf{G}(\mathbf{r},t) \\
    & + j\omega_{0}\epsilon_0\tilde{\mathbf{E}}(\mathbf{r},t)+\epsilon_0 \partial_t \tilde{\mathbf{E}}(\mathbf{r},t)
\end{split}
\end{equation}
\end{subequations}
and satisfy Gauss' electric and magnetic laws:
\begin{subequations}
\label{eq:maxwell}
\begin{equation}
    \nabla\cdot\epsilon_0 \tilde{\mathbf{E}}(\mathbf{r},t) = \rho_i (\vb{r}_s, t) +  \rho(\mathbf{r},t)\exp{-j\omega_{0}t},
\end{equation}
\begin{equation}
    \nabla\cdot\tilde{\mathbf{B}}(\mathbf{r},t) = 0.
\end{equation}
\end{subequations}
It is important to note at this point that $\mathbf{G}(\vb{r},t)$ and $\rho(\vb{r},t)$ are not necessarily narrowband. As a result, they cannot be decomposed into fast and slow varying  components. As usual, boundary conditions need to be imposed on $\tilde{\mathbf{E}}(\mathbf{r},t)$ and $\tilde{\mathbf{B}}(\mathbf{r},t)$ on sections of the outer boundary $\partial\Omega$. These are assumed to either Dirichlet, Neumann or impedance boundary conditions on non-overlapping surfaces $\partial\Omega_{D}$, $\partial\Omega_{N}$ and $\partial\Omega_{I}$, and are defined as follows:
\begin{subequations}
\begin{equation}
    \hat{n}\times\tilde{\mathbf{E}}(\mathbf{r},t) = \mathbf{\Psi}_{D}(\mathbf{r},t) \; \text{on}\; \Omega_{D},
\end{equation}
\begin{equation}
    \hat{n}\times\mu^{-1}\tilde{\mathbf{B}}(\mathbf{r},t) = \mathbf{\Psi}_{N}(\mathbf{r},t) \; \text{on}\; \Omega_{N},
\end{equation}
\begin{equation}
    \hat{n}\times\mu^{-1}\tilde{\mathbf{B}}(\mathbf{r},t) - 
    Y\hat{n}\times\hat{n}\times\tilde{\mathbf{E}}(\mathbf{r},t)= \mathbf{\Psi}_{I}(\mathbf{r},t) \; \text{on}\; \Omega_{I}.
\end{equation}
\end{subequations}
Note, it is assumed that $\partial\Omega_{D}+\partial\Omega_{N}+\partial\Omega_{I}=\partial\Omega$.
The evolution of the macroparticles in space and time is determined by solving for the equations of motion with the acceleration determined by the Lorentz force. This yields the following coupled system of equations for ordinary differential equations (ODEs) for  $\mathbf{v}_{p}(t)$ and $\mathbf{r}_{p}(t)$:
\begin{subequations}
\label{eq:lorentz}
\begin{equation}
    \frac{d\mathbf{v}_{p}(t)}{dt} = \frac{q}{m}\left[\tilde{\mathbf{E}}(\mathbf{r}_{p}(t),t)+\mathbf{v}_{p}(t)\times\tilde{\mathbf{B}}(\mathbf{r}_{p}(t),t)\right]\exp{j\omega_{0}t}
\end{equation}
\begin{equation}
    \frac{d\mathbf{r}_{p}(t)}{dt} = \mathbf{v}_{p}(t)
\end{equation}
\end{subequations}
In what follows, we present a method to self-consistently solve both Maxwell's equations and equations of motion, especially under the narrow band approximation. 

\subsection{Discretization in Space and Time}

Solutions to Eq. \eqref{eq:maxwell} are obtained by spatially representing $\tilde{\mathbf{E}}(\mathbf{r},t)$ and $\tilde{\mathbf{B}}(\mathbf{r},t)$ in terms of Whitney edge and face basis functions respectively:
\begin{equation}
    \begin{split}
     \tilde{\mathbf{E}}(\mathbf{r},t) &= \sum_{i=1}^{N_{e}} e_{i}(t)\mathbf{W}^{1}_{i}(\mathbf{r}), \\
     \tilde{\mathbf{B}}(\mathbf{r},t) &= \sum_{i=1}^{N_{f}} b_{i}(t)\mathbf{W}^{2}_{i}(\mathbf{r}).
    \end{split}
\end{equation}
Here, $\mathbf{W}^{1}_{i}(\mathbf{r})$ and $\mathbf{W}^{2}_{t}(\mathbf{r})$ represent the lowest order Whitney edge function defined on the $i$th edge and face function defined on the $t$th face, respectively. Further, $N_{e}$ and $N_{f}$ denote the number of edges and faces in the tetrahedral mesh used to discretize $\Omega$; details on mixed finite elements can be found \cite{Boss88,crawford2020unconditionally,Wong95,He06,Kett99} and reference therein. Using Gakerkin testing results in the following matrix ODE to solve for the vector of field coefficients $\bar{B}(t)$ and $\bar{E}(t)$ in time:
\begin{equation}\label{eq:maxwell_semi}
    \left[\Bar{\Bar{S}}\right]\cdot \mqty[\bar{B}(t) \\\bar{E}(t) ]
    + \left[\Bar{\Bar{M}}\right]\cdot \mqty[\partial_t \bar{B}(t)\\ \partial_t \bar{E}(t) ] = \bar{\bar{F}}
\end{equation}
where the various matrix definitions are as follows:
\begin{align}\label{eq:maxwell_mat_defs}
\begin{split}
     \left[\Bar{\Bar{S}}\right] &= \mqty[0&  [\curl] \\ -[\curl]^T & 0 ]+j\omega_{0}\mqty[ [\star_{\mu^{-1}}] & 0 \\ 0 &[\star_{\epsilon}]], \\
    \left[\Bar{\Bar{M}}\right] &= \mqty[ [\star_{\mu^{-1}}] & 0 \\ 0 &[\star_{\epsilon}]], \\
    \bar{\bar{F}} &= -\mqty[0\\ \frac{\bar{J_{i}}(t)}{\epsilon_0}]-\mqty[0\\ \frac{\partial_t\bar{G}(t)}{\epsilon_0}\exp{-j\omega_{0}t} ].
\end{split}
\end{align}
Furthermore, $ \bar{B} (t) = \left [ b_1(t), b_2(t), \cdots, b_{N_f} (t) \right ]^T$, $\bar{E}(t) = \left [ e_1(t), e_2 (t) , \cdots, e_{N_e}(t) \right ]^T$, $\bar{G}(t) = \left [ g_1(t), g_2 (t) , \cdots, g_{N_e}(t) \right ]^T$ where $g_{i}(t) = \left\langle \vb{W}^{(1)}_{i}, \mathbf{G}(\mathbf{r},t)\right\rangle$ and  $\bar{J}_{i}(t) = \left [ j_{1}(t), j_{2} (t) , \cdots, j_{N_e}(t) \right ]^T$ where $j_{j}(t) = \left\langle \vb{W}^{(1)}_{j}, \tilde{\vb{J}}_{i}(\mathbf{r}_{s},t)\right\rangle$. To convert \eqref{eq:maxwell_semi} into a discrete stencil at different intervals of time, we utilize a Newmark-$\beta$ \cite{zienkiewicz1977new} scheme with $\gamma=0.5$ and $\beta=0.25$. 
Furthermore, we define $\Delta_{t,\omega_{0}}=(30\xi f_{\text{max}})^{-1}$ and $\Delta_{t,\omega_{bw}}=(30\xi f_{\text{bw}})^{-1}$ as the timestep size appropriate for the fast varying and downshifted systems respectively, with $\xi$ being a real oversampling factor.
This choice of parameters results in representing the field coefficients in terms of second order Lagrange polynomials in time and testing with an average acceleration condition (represented by $W(t)$) such that for $t\in \left [t_{n-1},t_{n+1}\right]$ where $t_n = n \Delta_{t,\omega_{\text{bw}}}$
\begin{subequations}
	\begin{equation}
	\begin{split}
		\left(\begin{matrix}
			\bar{B}(t) \\
			\bar{E}(t)
		\end{matrix}\right) & = \sum\limits_{k = 0}^{2} N_{n,k}(t)\left(\begin{matrix}
			\bar{B}(t_{n+k-1}) \\
			\bar{E}(t_{n+k-1})
		\end{matrix}\right) \\
		& = \sum\limits_{k = 0}^{2} N_{n,k}(t) \left(\begin{matrix}
			\bar{B}^{n+k-1} \\
			\bar{E}^{n+k-1}
		\end{matrix}\right),
		\end{split}
	\end{equation}\\
	\begin{equation}\label{eq:newmark_L}
		L_{n,k}(t) = \prod_{\substack{j=0 \\ j\neq k}}^{2}
					\frac{t-t_{n+1-j}}{t_{n+1-k}-t_{n+1-j}} 
	\end{equation}\\
	\begin{equation}\label{eq:newmark_N}
		N_{n,k}(t) = \begin{cases}
		        L_{n,k}(t) & t \in \left[t_{n-1},t_{n+1}\right] \\
					0 & \text{otherwise}.
		 \end{cases}
	\end{equation}
\end{subequations}

\begin{equation}\label{eq:newmark_test_function}
	W_n(t) = 
	\begin{cases}
		\frac{t_n-t}{\Delta_t} & t \in \left[t_{n-1},t_{n}\right] \\
		\frac{t-t_n}{\Delta_t} & t \in \left[t_{n},t_{n+1}\right] \\
		0	& \textrm{otherwise}
	\end{cases}.
\end{equation}
 

\subsection{Discretization and Exact Charge Conservation \label{sec:ccons_metric}}

Next, we re-examine rubrics of a charge conserving PIC scheme from a slightly different perspective. The discrete divergence of the set of equations derived \eqref{eq:maxwell_semi} should yield Gauss' laws. But a word of caution, the discrete equations are a result of measuring their continuous counterparts in time. 
More precisely, we note that that the discrete divergence of Ampere's laws 
yields
\begin{equation}\label{eq:maxwell_semi_div}
\begin{split}
    \epsilon_{0}\left[\grad\right]^T \left(\left[\star_{\epsilon}\right]\partial_t\bar{E}(t)+j\omega_{0}\left[\star_{\epsilon}\right]\bar{E}(t)\right) & = - \left[\grad\right]^T\bigg ( \bar{J}_i (t)   \\
    & +  e^{-j\omega_{0}t} \partial_t\bar{G} (t) \bigg).
\end{split}
\end{equation}
Discretizing Gauss law, 
however, yields
\begin{equation}\label{eq:gaussDiscrete}
\epsilon_0 \left [ \grad \right ]^T  \left[\star_{\epsilon}\right] \bar{E}(t) = \bar{\rho}_i (t) + \bar{\rho}(t)
\end{equation}
where $\bar{\rho}_i(t) = [\rho_i (t), \cdots, \rho_{N_n}(t )]^T$, $\rho_i (t) = \langle W_i^0 (\vb{r}), \rho_i (\vb{r}_s,t) \rangle$, and $\bar{\rho}(t) = [\rho_1(t), \cdots, \rho_{N_n}(t)]^T = \left [ \grad \right ]^T \bar{G}(t)e^{-j \omega_0 t}$. Note, in the above equations, the $\left [ \grad \right ]^T $ matrix effects a divergence of the flux density (which lies in the dual grid) in terms of quantities defined on the primal grid. In practice, as one \emph{only} evolves the curl equations, the solution at every time step will automatically satisfy \eqref{eq:maxwell_semi_div} provided care is taken in constructing the right hand side \cite{Oconnor_time_integration}. But what is critical for satisfaction of Gauss' law is that this discrete solution  also satisfy \eqref{eq:gaussDiscrete}.  In the text that follows, we take some liberties in verbiage: (a) it will be \emph{understood} that \eqref{eq:maxwell_semi_div} and \eqref{eq:gaussDiscrete} are not explicitly discretized in time and evolved; (b) the system of equations resulting from taking the discrete divergence of the time discretized Ampere's laws and those from temporal discretization of \eqref{eq:maxwell_semi_div} will be identical; (c) we use the word discretization and measurement interchangeably they imply an inner product with a temporal basis function; and (d) our goal is to understand how one should measure or discretize \eqref{eq:gaussDiscrete} so as to to be consistent with the measurement of \eqref{eq:maxwell_semi_div}. Indeed, if we choose different temporal basis to measure 
\eqref{eq:maxwell_semi_div} and \eqref{eq:gaussDiscrete}, it must be shown that the two measures are consistent. For instance, if the basis functions used to measure \eqref{eq:maxwell_semi_div} are first order and those used to measure \eqref{eq:gaussDiscrete} are delta functions, they will yield different results; this is a point that we will return to later. Note, we have specifically, assumed that $\bar{G}(t)$ is known. Indeed, as shown in \cite{Oconnor_time_integration,crawford2021rubrics},  using $\bar{G}(t)$ is critical for exact satisfaction of Gauss' electric law. To understand these issues better, we analyze two cases; (a) $\omega_0 = 0$ and (b) $\omega_0 \neq 0$. To simplify our discussion, we assume both $\bar{J}_i (t) = 0 = \bar{\rho}_i (t)$. 

\subsubsection{Case 1: $\omega_{0}=0$}

When $\omega_{0}=0$, \eqref{eq:maxwell_semi_div} simplifies to
\begin{equation}\label{eq:maxwell_semi_div_omega_0}
    \epsilon_{0} \left[\grad\right]^T \left[\star_\epsilon \right]\partial_t\bar{E}(t) = -  \left[\grad\right]^T \partial_t \bar{G} (t)
\end{equation}
Note, parenthetically, we note that $\left[\grad\right]^T\bar{G}(t)=- \bar{\rho}(t)$. In keeping with the Newmark-$\beta$ scheme, to ensure late time stability, we have to use $W_n(t)$ as testing functions in time.  \emph{If a scheme is charge conserving, then } at any point $t_n$, the system arising from the discrete divergence of Ampere's law 
\begin{equation}\label{eq:ampere_omega_0}
    \langle W_n (t), \epsilon_{0} \left[\grad\right]^{T} \left[\star_\epsilon \right]\partial_t\bar{E}(t) \rangle = \langle W_n (t),  \partial_t \bar{\rho} (t) \rangle 
\end{equation}
should satisfy Gauss' law, but under what measure. To make the ensuing text notationally less dense, we use $\bar{\phi}(t) = \epsilon_0 \left [ \grad \right ]^{T}  \left[\star_{\epsilon}\right] \bar{E}(t)$. Thus, \eqref{eq:ampere_omega_0} can be written as
\begin{align}\label{eq:divAmpereTest}
    \langle W_n (t), \partial_t \bar{\phi}(t) \rangle & = \langle W_n (t), \partial_t \bar{\rho}(t) \rangle.
\end{align}
Evaluation of this inner product results in 
\begin{equation}\label{eq:ampere_stencil_omega_0}
\frac{\bar{\phi}^{n+1}-\bar{\phi}^{n-1}}{2}-\frac{\bar{\rho}^{n+1}-\bar{\rho}^{n-1}}{2}=0,
\end{equation}
which is the relations that  coefficients of the electric field will satisfy in keeping with the substitution defined earlier. 

The questions are two fold; (a) since Gauss' law in \eqref{eq:gaussDiscrete} is never solved, what is an equivalent discrete system, and (b) will the update equation for this system be consistent with those obtained from \eqref{eq:divAmpereTest}.   This may seem trivial as both sides of \eqref{eq:maxwell_semi_div_omega_0} have a time derivative and one can evaluate them \emph{analytically} to obtain \eqref{eq:gaussDiscrete} together with null initial conditions. 

But it is important to remember (a) that right hand side of Ampere's law  is deliberately chosen to be different from the conventional one where one discretizes current \cite{o2021set} and (b)    we effect the solution of \eqref{eq:maxwell_semi} by choosing temporal basis sets such that the solution is unconditionally stable, and do not treat it as a first order ODE. Under these circumstances, we can ask what should $\widetilde{W}_n(t)$ to reduce \eqref{eq:gaussDiscrete} to a discrete system. To make our analysis as general as possible, lets us denote $\widetilde{W}_{n}(t)$ as a basis that is used to discretize \eqref{eq:gaussDiscrete}:
\begin{equation}
\langle \widetilde{W}_n (t), \epsilon_0 \left [ \grad \right ]^{T}  \left[\star_{\epsilon}\right] \bar{E}(t) \rangle = \langle \widetilde{W}_n (t), \bar{\rho}(t) \rangle. 
\end{equation}
Using the abbreviated notation defined earlier, we get
\begin{equation}\label{eq:disc_gauss_omega_0}
    \langle \widetilde{W}_n (t),  \bar{\phi}(t) \rangle = \langle \widetilde{W}_n (t), \bar{\rho}(t) \rangle
\end{equation}
Consider two potential choices; for $\widetilde{W}_n (t) = \delta \left ( t - t_{n+1} \right ) $, one gets
\begin{equation}
\begin{split}
    \langle \delta (t - t_{n+1} ), \bar{\phi}(t) \rangle  & = \langle \delta (t - t_{n+1} ), \bar{\rho}(t)\rangle \\
    \bar{\phi}^{n+1}& = \bar{\rho}^{n+1}
    \end{split}
\end{equation}
and for $\widetilde{W}_n (t) = W_n  ( t )  $, one gets
\begin{equation}\label{eq:disc_gauss_stencil_omega_0}
\frac{\bar{\phi}^{n+1}+2\bar{\phi}^{n}+\bar{\phi}^{n-1}}{4}-\frac{\bar{\rho}^{n+1}+2\bar{\rho}^{n}+\bar{\rho}^{n-1}}{4}=0.
\end{equation}
In \eqref{eq:disc_gauss_stencil_omega_0}, if we were to assume an quiescent initial condition at $t = 0$, it is trivial to show that one recovers solutions obtained in \eqref{eq:ampere_stencil_omega_0}. However, for \eqref{eq:disc_gauss_stencil_omega_0} one needs Gauss' laws to be satisfied at both $t = \left \{ 0, \Delta_t\right \}$ for it to be consistent with solutions in \eqref{eq:ampere_stencil_omega_0}. Indeed, one can use \emph{any} test function as long as it is piecewise continuous, causal and one imposes additional initial conditions. 

We digress a little to note that this is a consequence of using $\vb{G}(t,\vb{r})$ and \emph{not} $\vb{J}(t,\vb{r})$. To set the stage, consider $\bar{J}(t) = \left [ \tilde{j}_1(t), \tilde{j}_2 (t) , \cdots, \tilde{j}_{N_e}(t) \right ]^T$ where $\tilde{j}_{i}(t) = \left\langle \vb{W}^{(1)}_{i}, \mathbf{J}(\mathbf{r},t)\right\rangle$. The standard discretized Ampere's yields
\begin{equation}
\begin{split}
    \epsilon_{0}\left[\grad\right]^T \left[\star_{\epsilon}\right]\partial_t\bar{E}(t) =   \left[\grad\right]^T\bar{J} (t).
\end{split}
\end{equation}
Using $\bar{\eta}(t) = \left[\grad\right]^T\bar{J} (t)$ and testing with $W_{n}(t)$ simplifies the notation to
\begin{equation}
    \left\langle W_{n}(t),\partial_{t} \bar{\phi}(t)\right\rangle = \left\langle W_{n}(t),\bar{\eta}(t)\right\rangle
\end{equation}
Evaluating the integrals as listed above yields the following stencil:
\begin{equation}\label{eq:tested_ampere_J}
    \bar{\phi}^{n+1}-\bar{\phi}^{n-1} = \Delta_{t,\omega_{\text{bw}}} \frac{\bar{\eta}^{n+1} + 2\bar{\eta}^{n} + \bar{\eta}^{n-1}}{4}
\end{equation}
But Gauss' law is slightly different; indeed, $\bar{\rho}(t) = \int_o^t d\tau\bar{\eta}(\tau)$. This implies that the discrete evolution of Gauss' law yields
\begin{equation}\label{eq:tested_gauss_omega_0}
    \left\langle W_{n}(t),\bar{\phi}(t)\right\rangle = \left\langle W_{n}(t),\bar{\rho}(t)\right\rangle
\end{equation}

Then \eqref{eq:tested_gauss_omega_0} simplifies to
\begin{equation}\label{eq:tested_gauss_omega_0_p2}
    \bar{\phi}^{n+1} + 2\bar{\phi}^{n} + \bar{\phi}^{n-1} = \bar{\rho}^{n+1} + 2\bar{\rho}^{n} + \bar{\rho}^{n-1}. 
\end{equation}
But as \emph{only} $\bar{\eta} (t)$ is available, one would need to integrate this to obtain the charge density. Choosing a backward Euler scheme for illustration, we obtain
\begin{equation}\label{eq:tested_gauss_omega_0_pointwise}
    \bar{\rho}^{n} = \bar{\rho}^{n-1} + \Delta_{t,\omega_{\text{bw}}} \bar{\eta}^{n}.
\end{equation}
Using \eqref{eq:tested_gauss_omega_0_pointwise} in \eqref{eq:tested_ampere_J} results in 
\begin{equation}\label{eq:tested_ampere_J_rho}
    \bar{\phi}^{n+1}-\bar{\phi}^{n-1} = \frac{1}{4}\left ( \bar{\rho}^{n+1} + \bar{\rho}^{n} - \bar{\rho}^{n-1} - \bar{\rho}^{n-2} \right )
\end{equation}
It is apparent from above equations, that irrespective of the integration scheme, the coefficient obtained by solving Ampere's law will not satisfy the discrete Gauss' law.

The above discussion presents the nuances of the challenge in ensuring that the solution of the discrete curl equations satisfy Gauss' law. When $\omega_0 = 0$, one can somewhat get around the challenge by defining $\bar{G}(t)$. As we will see, this challenge cannot be overcome when $\omega_0 \neq 0$. 

\subsubsection{Case 2: $\omega_{0} \neq 0$}
Next, we consider the effect of adding an analytically prescribed fast-varying function to our field solution.
As before, we begin our analysis with the divergence of Ampere's Law in \eqref{eq:maxwell_semi_div} and obtain
\begin{equation}
\epsilon_{0}\left[\grad\right]^T \left(\left[\star_{\epsilon}\right]\partial_t\bar{E}(t)+j\omega_{0}\left[\star_{\epsilon}\right]\bar{E}(t)\right) = - \left[\grad\right]^T\left(  e^{-j\omega_{0}t} \partial_t\bar{G} (t) \right). 
\end{equation}
Using the notation defined earlier, and testing with $W_{n}(t)$, we get
\begin{equation}
    \left\langle W_{n}(t), \partial_{t}\bar{\phi}(t) + j\omega_{0}\bar{\phi}(t)\right\rangle = \left\langle W_{n}(t), \exp{-j\omega_{0}t} \partial_{t}\bar{\rho}(t)\right\rangle
\end{equation}.
Evaluating the inner product integrals, yields
\begin{equation}\label{eq:ampere_stencil_omega_neq_0}
\begin{split}
&\left(j\omega_{0}\frac{\Delta_{t,\omega_{\text{bw}}}}{4} + \frac{1}{2}\right)\bar{\phi}^{n+1} = - j\omega_{0}\frac{\Delta_{t,\omega_{\text{bw}}}}{2}\bar{\phi}^{n} - \left(j\omega_{0}\frac{\Delta_{t,\omega_{\text{bw}}}}{4}-\frac{1}{2}\right)\bar{\phi}^{n-1} \\
&+ \beta^{n+1}(\omega_{0})\bar{\rho}^{n+1} + \beta^{n}(\omega_{0})\bar{\rho}^{n} + \beta^{n-1}(\omega_{0})\bar{\rho}^{n-1}
\end{split}
\end{equation}
where $\beta^{n+1}(\omega_{0})$, $\beta^{n}(\omega_{0})$ and $\beta^{n-1}(\omega_{0})$ are defined in  Section \ref{sec:Wn_innerprods}.
If charge conservation is achieved, then we should be able to find a testing function $\widetilde{W}_{n}(t)$ that, when used to measure Gauss' Law will reduce it to a form that is satisfied by the solution to \eqref{eq:ampere_stencil_omega_neq_0}.

Let us first begin by setting $\widetilde{W}_{n}(t)=\delta(t-t_{n})$. Testing Gauss' law with this function will lead to
\begin{equation}
    \left\langle \delta(t-t_{n}), \bar{\phi}(t)\right\rangle = \left\langle \delta(t-t_{n}), \exp{-j\omega_{0}t} \bar{\rho}(t)\right\rangle
\end{equation}
which, when evaluated leads to the following relation for $\bar{\phi}^{n}$
\begin{equation}
    \bar{\phi}^{n} = \exp{-j\omega_{0}t_{n}}\bar{\rho}^{n}
\end{equation}
Comparing this relation to \eqref{eq:ampere_stencil_omega_neq_0}, we find that the pointwise relation for $\phi_{n}$ does not satisfy the stencil for the divergence of Ampere's Law. \emph{Interestingly, the solutions do agree in the limit when $\omega_{0}\rightarrow 0$, but otherwise, they are inconsistent.}
Likewise, if we were to perform a similar analysis with $\widetilde{W}_{n}(t) = W_{n}(t)$, we end up with the following update expression for $\phi^{n+1}$
\begin{equation}\label{eq:gauss_stencil_W_omega_neq_0}
\begin{split}
\frac{\Delta_{t,\omega_{\text{bw}}}}{4}\bar{\phi}^{n+1} =& - \frac{\Delta_{t,\omega_{\text{bw}}}}{2}\bar{\phi}^{n} - \frac{\Delta_{t,\omega_{\text{bw}}}}{4}\bar{\phi}^{n-1}
+ \alpha^{n+1}(\omega_{0})\bar{\rho}^{n+1} \\
&+ \alpha^{n}(\omega_{0})\bar{\rho}^{n} + \alpha^{n-1}(\omega_{0})\bar{\rho}^{n-1}
\end{split}
\end{equation}
Once again, the coefficients $\alpha^{n+1}(\omega_{0})$, $\alpha^{n}(\omega_{0})$ and $\alpha^{n-1}(\omega_{0})$ refer respectively to the inner products of $W_{n}(t)$ and $\exp{-j\omega_{0}t}$ multiplying the basis functions used to represent the field quantities in time $N_{n,0}(t)$, $N_{n,1}(t)$and $N_{n,2}(t)$ respectively (defined in \eqref{eq:newmark_N}). As before, \eqref{eq:gauss_stencil_W_omega_neq_0} and \eqref{eq:ampere_stencil_omega_neq_0} agree in the limit of $\omega_{0}\rightarrow 0$, they are inconsistent for finite $\omega_{0}$. 

While neither choice of $\widetilde{W}_{n}(t)$ result in discrete systems that are consistent with \eqref{eq:ampere_stencil_omega_neq_0}, and there may never be, we take a different and more robust path. In addition to the above difficulties, we note the following: The standard Newmark time stepping scheme excites a null space that is time independent. While the null space can be small, it will not have a trivial divergence and will corrupt the satisfaction of Gauss' law \cite{crawford2020unconditionally}.  A way to overcome both problems is to use a Helmholtz decomposition or impose a Coulomb gauge in the discrete setting. The means to do so is elaborated next. 

\subsection{Quasi Helmholtz or Coulomb Gauge \label{sec:quasihelmholtz}}

Much of the development of what follows has been detailed in our earlier paper \cite{oconnor2021quasihelmholtz}. The following discussion is purely for completeness of the paper. In a nutshell, $\tilde{\mathbf{E}}(\mathbf{r},t)$ and $\tilde{\mathbf{B}}(\mathbf{r},t)$ can decomposed into solenoidal (components that are divergence free) and non-solenoidal components (have a finite curl). Hence, the use of the prefix ``quasi.'' In this framework, Gauss' laws are explicitly discretized and solved. Furthermore, while the solution for the solenoidal component has a null space, it divergence is \emph{exactly} zero. As a result, Gauss' laws are exactly satisfied. 

In what follows, we use $\bar{E}^{n}_{ns}$  refers to the non-solenoidal coefficients of the electric field at the $n$th timestep. Likewise, $\bar{E}^{n}_{s}$ and $\bar{B}^{n}_{s}$ refer  to the divergence-free solenoidal coefficients of the electric field and magnetic flux density, respectively. A complete prescription of all sub-matrices involved are provided in Section \ref{sec:qhmats}.

\subsubsection{Requisite Proejctors}

To project the non-solenoidal components of from the basis used for representing the electric field, we define projectors $[\bar{P}]^{\Sigma}_{e}$ and $[\bar{P}]^{\Lambda}_{e}$ that, when operated on the complete electric field, respectively extract the non-solenoidal and solenoidal components (where $\dagger$ represents a Moore-Penrose pseudoinverse):
\begin{subequations}
\begin{align}
    [\bar{P}]_e^{\Sigma} & = \Sigma(\Sigma^T \Sigma)^{\dagger} \Sigma^{T} \label{eq:proj_es}\\
    [\bar{P}]^{\Lambda}_e & = \mathcal{I} - [\bar{P}]_e^{\Sigma} \label{eq:proj_el}
\end{align}
\begin{equation}
\label{eq:bfld_projector}
\left [ \bar{P}\right ]^\Lambda_b = \mathcal{I} - \Sigma_m\left ( \Sigma_m^T \Sigma_m \right )^{\dagger} \Sigma_m^T 
\end{equation}
\end{subequations}
where $[\Sigma]=\epsilon_{0}[\bar{M}_{g}]$ and $[\Sigma]_m=[\div]^T$. Using these projectors,, we can now define a decomposition for the electric flux density as
\begin{equation}
\label{eq:efld_projected}
\bar{D}^n = \Sigma \bar{E}^n_{ns} + \left [ \bar{P}\right ] ^\Lambda_e \bar{D}^n
\end{equation}
and the magnetic flux density as
\begin{equation}
\label{eq:bfld_projected}
\bar{B}_{s}^n = \left [ \bar{P}\right ] ^\Lambda_b \bar{B}^n
\end{equation}
Using this projector, it is rather straightforward to show that the divergence of $\bar{B}^n$ is zero.

\subsubsection{Discrete System}

To use this decomposition, we use \emph{all} of Maxwell's equations. First given the decomposition, the application of the discrete divergence on \eqref{eq:bfld_projected} and \eqref{eq:efld_projected}  results in zero and 
\begin{equation}
\label{eq:ns_equation}
\left[C_{z}^{e}\right]^{T}\left[\nabla\right]^{T}\left[\star_{e}\right]\left[\nabla\right]\left[C_{z}^{e}\right]\bar{E}_{ns}^{n}
=-\left[C_{z}^{e}\right]^{T}\left[\nabla\right]^{T}e^{-j\omega_{0} t}\bar{G}^{n}
\end{equation}
Likewise, using decomposition in \eqref{eq:maxwell_semi} and with \eqref{eq:ns_equation} results in 
\begin{equation}
    \label{eq:sol_disc}
    \begin{split}
        \left[\bar{Z}\right]_{11}\partial_{t}\bar{B}_{s}^{n}
        +& j\omega_{0}\left[\bar{Z}\right]_{11}\bar{B}_{s}^{n}
        +\left[\bar{Z}\right]_{12}\bar{E}_{s}^{n}=
        -\left[\bar{Z}\right]_{13}\bar{E}_{ns}^{n}\\
        \left[\bar{Z}\right]_{21}\partial_{t}\bar{E}_{s}^{n}
        + & j\omega_{0}\left[\bar{Z}\right]_{21}\bar{E}_{s}^{n}-
        \left[\bar{Z}\right]_{22}\bar{B}_{s}^{n}
        =-e^{-j\omega_{0}t}\partial_{t}\bar{G}^{n}\\
        &-\left[\bar{Z}\right]_{23}j\omega_{0}\bar{E}_{ns}^{n} -\left[\bar{Z}\right]_{23}\partial_{t}\bar{E}_{ns}^{n}
    \end{split}
\end{equation}
where $\bar{E}^{n}_{s}$ and $\bar{B}^{n}_{s}$ are extracted from the simply connected mesh using tree-cotree maps $\left[C^{e}_{c}\right]$ and $\left[C^{b}_{c}\right]$ that are defined in Section \ref{sec:qhmats}. The evaluation of time derivatives is effected through the basis defined earlier; it has been extensively discussed in \cite{Oconnor_time_integration}. 

As alluded to earlier, since we directly discretize Gauss' laws and since the discrete divergence of \eqref{eq:sol_disc} is identical to zero, charge conservation is exactly satisfied.
\subsection{Particle Push and Integration Scheme for $\bar{G}$ \label{sec:G_integ}}
Next, we describe the framework used to evolve particle trajectories following \eqref{eq:lorentz}.
We recall from Section \ref{sec:quasihelmholtz} that the error in the non-solenoidal component of the electric field is tied intimately to the accuracy in the evaluation of $\bar{G}(t)$, making it important to evaluate the particle positions and velocities to high accuracy. But it is also known that the Fourier transform of the velocity or the position of any particle is not necessarily narrow-band. As a result, the equations of motion and the path integral need to be evaluated at a step size $\Delta_{t,\omega_{0}}$, while samples of the field are only known at $\Delta_{t,\omega_{\text{bw}}}$.
There are two ways to get around this mismatch. First, one can incorporate the analytically known fast varying field components into the Lorentz update
\begin{equation}
\begin{split}
    \frac{d\mathbf{v}}{dt}(\mathbf{r}_{p},t) = \frac{q_{p}}{m_{p}} \mathbf{Re}\left\{\tilde{\mathbf{E}}(\mathbf{r}_{p},t)\exp{j\omega_{0}t}\right\} \\
    + \frac{q_{p}}{m_{p}}\mathbf{v}(\mathbf{r}_{p},t)\times \mathbf{Re}\left\{\tilde{\mathbf{B}}(\mathbf{r}_{p},t)\exp{j\omega_{0}t}\right\},
\end{split}
\end{equation}
which can be numerically integrated \emph{with the fast varying components treated analytically}. The specifics of the resulting integration rule can be written as
\begin{equation}
\begin{split}
\mathbf{v}(\mathbf{r}_{p}^{n+1},t_{n+1})-\mathbf{v}(\mathbf{r}_{p}^{n},t_{n}) = \\ \frac{q_{p}}{m_{p}} \sum_{k=0}^{2}\mathbf{E}(\mathbf{r}_{p}^{n-1-k},t_{n-1-k})\left(M^{r}_{n,k} + M^{i}_{n,k}\right) \\
+ \frac{q_{p}}{m_{p}}\sum_{k=0}^{2}\mathbf{v}(\mathbf{r}_{p}^{n-1-k},t_{n-1-k})\times\mathbf{B}(\mathbf{r}_{p}^{n-1-k},t_{n,1,k}) \\
\left(L^{r}_{n,k}+L^{i}_{n,k}\right)
\end{split}
\end{equation}
\begin{subequations}
where the operators used are defined as follows:
\begin{equation}
    L^{r}_{n,j,k} = \intop_{t_{n}}^{t_{n+1}} N_{n,j}(t)N_{n,k}(t) \cos (\omega_{0}t) dt,
\end{equation}
\begin{equation}
    L^{i}_{n,j,k} = - \intop_{t_{n}}^{t_{n+1}} N_{n,j}(t)N_{n,k}(t) \sin (\omega_{0}t) dt,
\end{equation}
\begin{equation}
    M^{r}_{n,j} = \intop_{t_{n}}^{t_{n+1}} N_{n,j}(t)\cos (\omega_{0}t) dt,
\end{equation}
\begin{equation}
    M^{i}_{n,j} = - \intop_{t_{n}}^{t_{n+1}} N_{n,j}(t)\sin (\omega_{0}t) dt.
\end{equation}
\end{subequations}
The particle positions can then be integrated consistently in time at $\Delta_{t,\omega_{\text{0}}}$ through a fourth order Adams-Bashforth stencil:
\begin{equation}\label{eq:newtons}
\begin{split}
\mathbf{r}_p^{n+1} &= \mathbf{r}_p^{n} + \frac{\Delta_{t,\omega_{0}}}{24}(55 \mathbf{v}_p^n - 59 \mathbf{v}_p^{n-1} + 37 \mathbf{v}_p^{n-2}-9\mathbf{v}_p^{n-3})
\end{split}
\end{equation}
Since the particle positions cannot be downshifted in the same way as the velocity update, some oversampling is still required to maintain accuracy, but as we show in Sec. \ref{sec:error_ens}, this factor is relatively small.
The computed trajectories can then be used to obtain $\bar{G}(t)$
\begin{equation}
\label{eq:G_integral}
\begin{split}
\bar{G}^{n}_{l}= \left\langle \mathbf{W}^{(1)}_{l}(\mathbf{r}_{p}(n\Delta_{t,\omega_{\text{bw}}})), \sum_{p=1}^{N_{p}} q \intop_{\mathbf{r}_{p}(0)}^{\mathbf{r}_{p}(n\Delta_{t,\omega_{\text{bw}}})}d\widetilde{\mathbf{r}}S(\mathbf{r} - \widetilde{\mathbf{r}})\right\rangle,
\end{split}
\end{equation}
where $n$ refers to the number of timesteps evolved by the EM system, and $l$ refers to a specific edge in the EM system.

In systems where the number of particles is not very large, and therefore the cost of a particle update is negligible in relation to the cost of a field solve, we can follow a simpler integration setup and simply interpolate the electric field to time points that are spaced at $\Delta_{t,\omega_{0}}$.
In essence, the magnitudes of the electric fields and magnetic flux density at the location of a given particle $p$, can be reconstructed from the downshifted quantities at a given time $t\in[t_{n-1},t_{n+1}]$ as
\begin{subequations}
\begin{align}
\mathbf{E}(\mathbf{r}_{p},t) = \sum_{i=1}^{N_{e}} \sum_{k=0}^{2}N_{n,k}(t)\bar{E}^{k}_{i} \exp{j\omega_{0} t_{n-1+k}} \mathbf{W}^{1}_{i}(\mathbf{r}_{p}) \\
\mathbf{B}(\mathbf{r}_{p},t) = \sum_{i=1}^{N_{f}}\sum_{k=0}^{2}N_{n,k}(t) \bar{B}_{i}^{k}\exp{j\omega_{0} t_{n-1+k}} \mathbf{W}^{2}_{i}(\mathbf{r}_{p})
\end{align}
\end{subequations}
These fields can then be used in a Lorentz update at the smaller stepsize.
Since the particle paths are now known at the finer timestep size $\Delta_{t,\omega_{0}}$, the path integral in \eqref{eq:G_integral} can be evaluated as
\begin{equation}
\intop_{\mathbf{r}_{p}(0)}^{\mathbf{r}_{p}(n\Delta_{t,\omega_{\text{bw}}})}d\widetilde{\mathbf{r}}S(\mathbf{r} - \widetilde{\mathbf{r}}) = \sum_{j=0}^{M-1} \intop_{\mathbf{r}_{p}(j\Delta_{t,\omega_{0}})}^{\mathbf{r}_{p}((j+1)\Delta_{t,\omega_{0}})}d\widetilde{\mathbf{r}}S(\mathbf{r} - \widetilde{\mathbf{r}}).  
\end{equation}
where $M$ refers to the number of timesteps needed at $\Delta_{t,\omega_{0}}$ to advance to a time of $n\Delta_{t,\omega_{\text{bw}}}$.
We show, once again, in Section \ref{sec:error_ens} that this approach can be used to get precision similar to the actively downshifted integration scheme described earlier in the section, but the rate of oversampling required is larger. Note, in both these methods, we have assumed non-relativistic motion. Studies are underway on efficient methods in relativistic regimes.

\section{Numerical Experiments \label{sec:results}}
In this section, we detail results obtained from an implementation of the ET-FEMPIC scheme described above. The results will be structured as follows: (A) First, we demonstrate the viability and computational gains of using envelope tracking in a linear EM system without particles by analysing the radiated power due to a monopole antenna. (B) Next. we demonstrate the accuracy of  the integration scheme developed in Section \ref{sec:G_integ}. Then, we proceed to analyse two systems with particles, thereby including non-linear effects from the active media. 
As a prelude, we note that the modulated gaussian functions used in some of the numerical examples are defined as follows:
\begin{equation}\label{eq:modgauss}
\begin{split}
    v(t) &= \cos(\omega_{0} t)\exp\left(-\frac{(t-6\sigma)^{2}}{2\sigma^{2}}\right) \\
    \sigma &= \frac{2}{\omega_{\text{bw}}}
\end{split}
\end{equation}

\subsection{Radiated Power from a Monopole Antenna \label{sec:monopole}}
\begin{figure}
\begin{subfigure}{0.5\textwidth}
    \centering
    \includegraphics[width=0.8\linewidth]{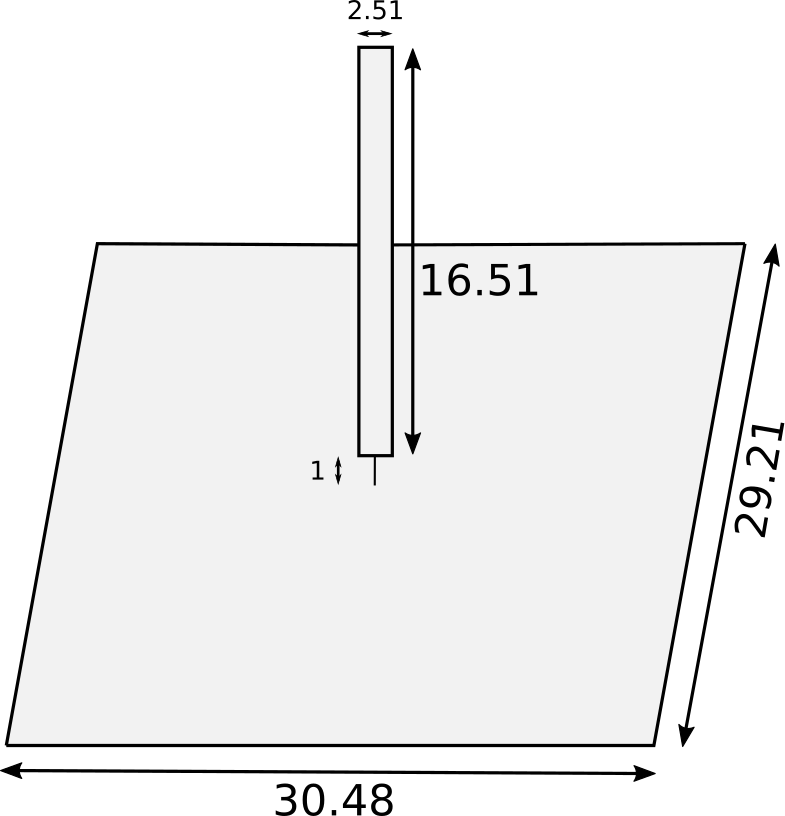}
    \caption{Geometry of the monopole antenna. The dimensions are in cm.}
    \label{fig:kong_monopole}
\end{subfigure}
\begin{subfigure}{0.5\textwidth}
    \centering
    \includegraphics[width=\linewidth]{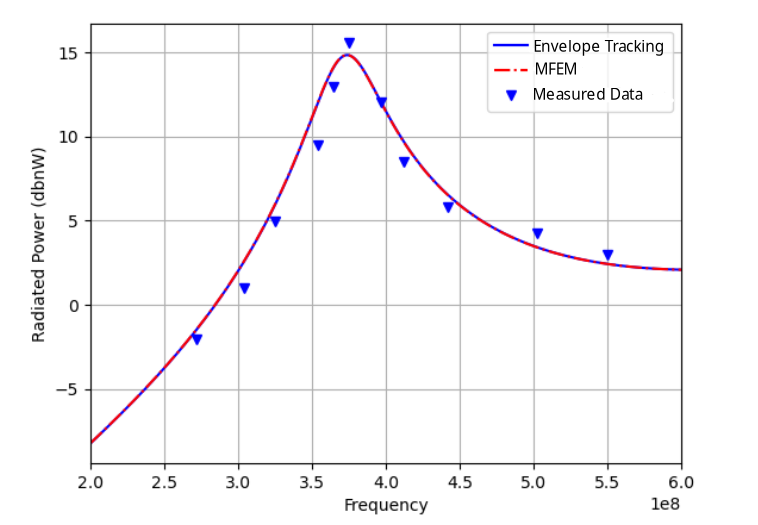}
    \caption{Power radiated due to a 1 mV source compared to measured data and regular MFEM}
    \label{fig:kong_power}
\end{subfigure}
\end{figure}
We consider a conducting strip suspended over a finite ground plane, as specified in Fig. \ref{fig:kong_monopole}.
The coupling between the EM system and the driving circuit is achieved across a vertical 1 cm edge going from the conducting plane to the strip.
The current driving the antenna is generated by a Thevenin source characterized by $f_{0}=300$ MHz and $f_{\text{bw}}=200$ MHz connected in series to a $100\ \Omega$ resistor.
The circuit subsystem itself was modelled through a Modified Nodal Analysis network \cite{mna}, with the constituent equations appropriately modified to account for the analytically known fast field components.
The voltage and current across the port feed were then used to compute the complex impedance of the feed as a function frequency. 
The radiated power curves for the antenna were computed from the impedance obtained from a regular MFEM solve and through the envelope tracking technique described in this work.
As can be seen in Fig. \ref{fig:kong_power}, there is good agreement in the radiated power as a function of frequency between a regular MFEM and the envelope tracking method, despite the timestep size in the latter being 2.5 times larger.

\subsection{Fidelity of the Particle Push Routine \label{sec:error_ens}}
Since the particle positions and velocities are not known to conform to the same frequency spread as the fields, we proposed two different time-stepping schemes in Section \ref{sec:G_integ} to accurately evolve Newton's equations.
The numerical results presented in this section demonstrate the viability of both methods. To set the stage, we consider a system consisting of one particle moving under the influence of $\mathbf{E}(\mathbf{r},t)=\hat{x} v(t)$ and $\mathbf{B}(\mathbf{r},t)=\hat{z} v(t)$ with $v(t)$ as defined in \eqref{eq:modgauss}.
The positions of the fields and velocities are then solved for using two different methods. First, we use a particle solver that operates at a smaller timestep size $\Delta_{t,\omega_{\text{max}}}$ than the field solve. The values of the field are then interpolated to the finer timesteps and used to evolve Newton's equations. Since the fields conform to the bandwidth amenable for an envelope tracking analysis, the interpolation should be as accurate as a time-marching routine that evolves at steps of $\Delta_{t,\omega_{\text{max}}}$. Second, we evolve the particle using the downshifted integration scheme described in Sec. \ref{sec:G_integ}. The final position curves obtained from both methods was compared against a solution obtained using a fourth order Adams Bashforth method evolved at $\Delta_{t,\text{bench}}=\Delta_{t,\omega_{\text{max}}}/10$. The relative error of the two methods in relation to this benchmark is reported in Table \ref{tab:err_G}. As is evident, both methods yield the same order of error over a range of center frequency/bandwidth combinations. 

\begin{table}
\begin{centering}
\begin{tabular}{|c|c|c|c|c|c|}
\hline
$f_{\text{bw}}$ (MHz) &
$N_{\text{Regular}}$ &
$\epsilon^{\mathcal{L}^{2}}\left(N_{\text{Regular}}\right)$ &
$N_{\text{Downshift}}$ &
$\epsilon^{\mathcal{L}^{2}}\left(N_{\text{Downshift}}\right)$\tabularnewline
\hline
\hline
20 &
100 &
$4.135\times10^{-8}$ &
21 &
$3.916\times10^{-8}$\tabularnewline
\hline
10 &
200 &
$3.308\times10^{-8}$ &
54 &
$6.392\times10^{-8}$\tabularnewline
\hline
5 &
400 &
$3.591\times10^{-8}$ &
117 &
$7.997\times10^{-8}$\tabularnewline
\hline
1 &
2000 &
$1.195\times10^{-8}$ &
501 &
$2.410\times10^{-8}$\tabularnewline
\hline
\end{tabular}
\par\end{centering}
\caption{Errors for the two particle integration methods, compared against numerically obtained data at $\Delta_{t,\text{bench}}=\Delta_{t,\omega_{\text{max}}}/10$. $N_{\text{Regular}}$ and $N_{\text{Downshift}}$ refer to the oversampling factor for the naively oversampled and downshifted methods respectively, and $\epsilon^{\mathcal{L}_{2}}$ refers to the $\mathcal{L}^{2}$ error for a given method compared against the benchmark data. We note that the oversampling factor required for the downshifted method is significantly lower than with naive oversampling. In each case, $f_{0} = 2$ GHz.}
\label{tab:err_G}
\end{table}

\begin{table*}[t]
    \centering
\begin{tabular}{|c|c|c|c|c|c|c|}
\hline
$t_{\text{max}}$(ns) &
$N_{t,\text{ET-MFEM}}$ &
$Q_{\text{ET-MFEM}}$ &
$N_{t,\text{MFEM}}$ &
$Q_{\text{MFEM}}$ &
$Q_{\text{EM-FEMPIC}}$ &
$Q_{\text{ET-FEMPIC}}$\tabularnewline
\hline
\hline
$135.6$ &
$1500$ &
$1079.1$ &
$13500$ &
$1084$ &
$903.7$ &
$899.4$\tabularnewline
\hline
$90.4$ &
$1000$ &
$556.4$ &
$9000$ &
$561.4$ &
$484.9$ &
$488.4$\tabularnewline
\hline
$67.8$ &
$750$ &
$231.4$ &
$6750$ &
$246.7$ &
$211.5$ &
$207.8$\tabularnewline
\hline
\end{tabular}
    \caption{Tabulated quality factor data in the Klystron solve. $N_{t}$ and $Q$ refer respectively to the number of timesteps used in the analysis and the respective quality factor obtained. As can be seen in the table, the Envelope tracking approach closely tracks the results predicted by the MFEM solve.}
    \label{tab:Q_klys}
\end{table*}

\subsection{Klystron exicted by a gap voltage \label{sec:klystron}}
\begin{figure}
    \centering
    \includegraphics[width=\linewidth]{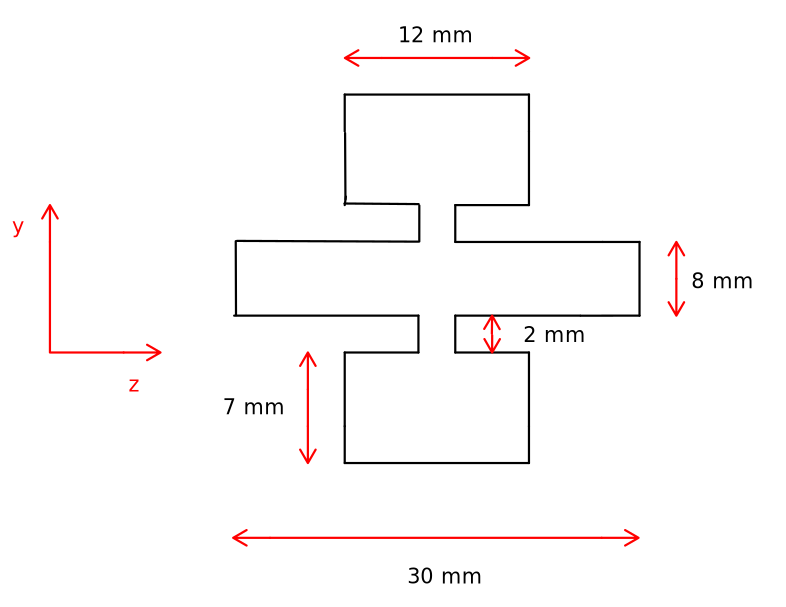}
    \caption{Schematic of the Klystron. The figure represents a cross-section in the $y$-$z$ plane. The device is assumed to be cylindrically symmetric about the $z$ axis. Furthermore, all walls are PEC.}
    \label{fig:klyst_geom}
\end{figure}
\begin{figure}
    \centering
    \includegraphics[width=\linewidth]{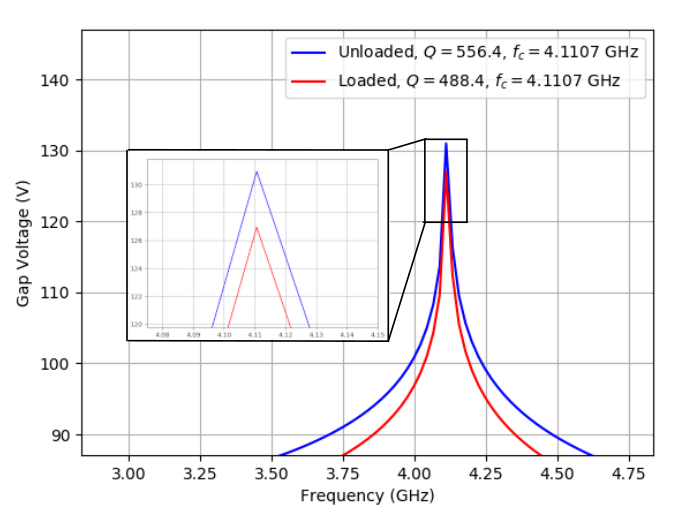}
    \caption{Gap voltage vs frequency in the loaded and unloaded case -- both run in the downshifted frame. We note that the peaks line up at the same frequency and the loaded response shows a dip in the saturated gap voltage thereby reducing the quality factor.}
    \label{fig:klyst_hilb}
\end{figure}
\begin{figure}
    \centering
    \includegraphics[width=\linewidth]{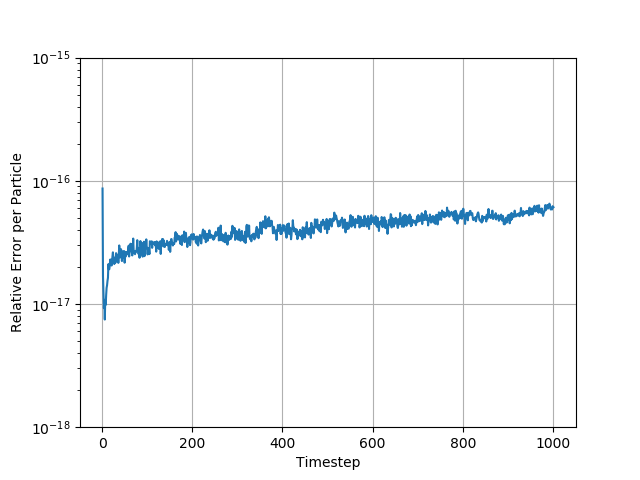}
    \caption{Plot of relative error per particle in satisfaction of discrete Gauss' Law for the Klystron setup simulated through ET-FEMPIC.}
    \label{fig:klyst_DGL}
\end{figure}
\begin{figure}
    \centering
    \includegraphics[width=\linewidth]{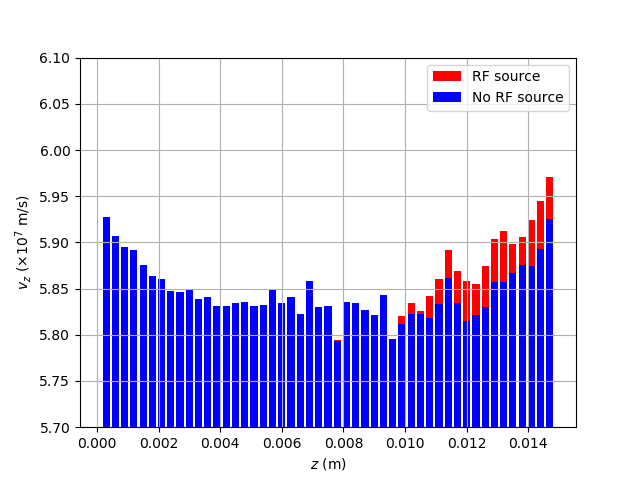}
    \caption{Velocity histogram at $t=100$ ns along the length of the Klystron showing particle acceleration when exposed to the RF source.}
    \label{fig:klyst_hist}
\end{figure}
\begin{figure}
    \centering
    \includegraphics[width=\linewidth]{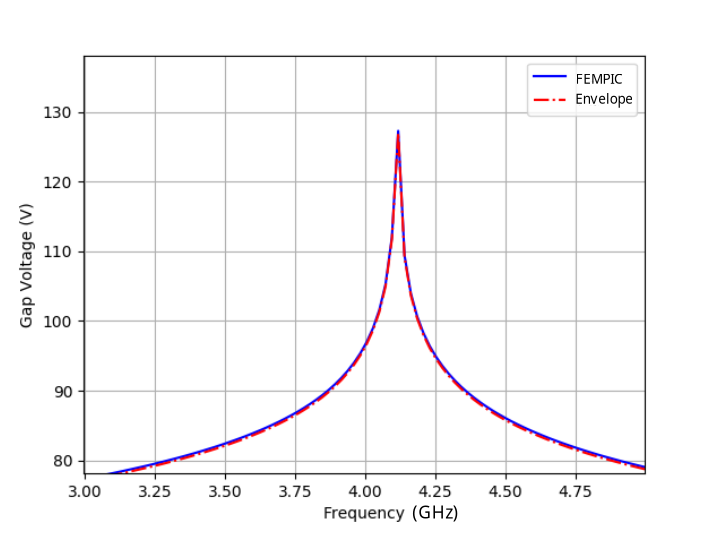}
    \caption{Comparison between the loaded cavity responses obtained through the regular and downshifted frames.}
    \label{fig:klyst_comp}
\end{figure}
Next, we analyze the EM response for a device with strong particle effects.
Specifically, we examine the reduction in the quality factor of a Klystron under beam loading, with the set-up  as shown in Fig. \ref{fig:klyst_geom}.
A sinusoidal RF source placed at the connecting neck, a 40 kV, 3A beam was introduced on one end of the feed-tube. The gap voltage as a function of frequency was then computed from the Fourier transform of the measured electric field across an edge spanning the walls of the neck.
The quality factor of the cavity was estimated by locating the half power points beneath the resonance peak, and computing $Q$ as
\begin{equation}
    Q = \frac{f_{\text{peak}}}{\Delta f}
\end{equation}
We note from Fig. \ref{fig:klyst_hilb} and Table. \ref{tab:Q_klys} that the quality factor when the klystron is loaded drops slightly in comparison to an unloaded run, with some of the energy of the cavity used to accelerate the particles. 
This acceleration can be seen by observing a histogram of particle velocities as a function of position along the length of the tube, as shown in Fig. \ref{fig:klyst_hist}.
Furthermore, the quality factor derived using ET-FEMPIC are close to those derived from a traditional EM-FEMPIC solve \emph{but computed at approximately the tenth the number of time steps}. Furthermore, the drop observed is similar to that observed in \cite{wilsen2002klystron} simulated using an axi-symmetric finite difference time domain code. Finally, we note from Fig. \ref{fig:klyst_DGL} that Gauss' Law is satisfied to machine tolerance. 

\subsection{Cascaded Klystron \label{sec:d_klystron}}

\begin{figure}
    \centering
    \includegraphics[width=\linewidth]{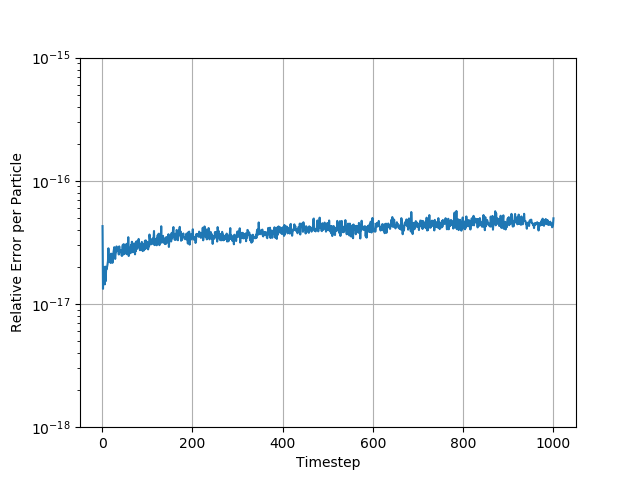}
    \caption{Plot of relative error per particle in satisfaction of discrete Gauss' Law for the cascaded Klystron system simulated through ET-FEMPIC.}
    \label{fig:doub_klyst_DGL}
\end{figure}
\begin{table*}[t]
\begin{centering}
\begin{tabular}{|c|c|c|c|}
\hline 
Distance between ports (cm) &
$\left|\mathbf{e}_{1}\right|$ &
$\left|\mathbf{e}_{2}\right|$ &
$\left|\mathbf{e}_{2}^{\text{2 Ports excited}}\right|$\tabularnewline
\hline 
\hline 
$15$ &
$131.4$ &
$15.31$ &
$143.5$\tabularnewline
\hline 
$30$ &
$134.1$ &
$1.21$ &
$133.7$\tabularnewline
\hline 
\end{tabular}
\par\end{centering}
\caption{Frequency Domain results at the feed locations for the double Klystron. For each separation distance, a unit impulse was imposed on the first port and the response measured on the second port. We note that at 15 cm, the magnitude of the field seen at the second port is markedly higher than at 30 cm.}
\label{tab:klsytron_FD}
\end{table*}
\begin{table*}
\begin{centering}
\begin{tabular}{|c|c|c|c|c|c|c|}
\hline 
$t_{\text{max}}$(ns) &
$N_{t,\text{EM-FEMPIC}}$ &
$Q_{\text{EM-FEMPIC},\text{P1}}$ &
$Q_{\text{EM-FEMPIC},\text{P2}}$ &
$N_{t,\text{ET-FEMPIC}}$ &
$Q_{\text{ET-FEMPIC},\text{P1}}$ &
$Q_{\text{ET-FEMPIC},\text{P2}}$\tabularnewline
\hline 
\hline 
$135.6$ &
$13500$ &
$1081.9$ &
$1086.1$ &
$1500$ &
$1075.1$ &
$1060.1$\tabularnewline
\hline 
$90.4$ &
$9000$ &
$503.4$ &
$499.4$ &
$1000$ &
$500.1$ &
$491.1$\tabularnewline
\hline 
$67.8$ &
$6750$ &
$247.3$ &
$248.1$ &
$750$ &
$246.1$ &
$235.1$\tabularnewline
\hline 
\end{tabular}
\par\end{centering}
 \caption{Tabulated quality factor data in the Double Klystron solve. All reported values are for the beam-loaded setup. $N_{t}$ and $Q$ refer respectively to the number of timesteps used in the analysis and the respective quality factor obtained. Likewise, quantities with subscripts $P1$ and $P2$ respectively referred to measurements made at ports 1 and 2 respectively. As can be seen in the results, the Envelope tracking approach closely tracks the results predicted by the MFEM solve.}
\label{tab:klystron_data}
\end{table*}
Finally, in order to demonstrate the efficacy of ET-FEMPIC on a complex, larger scale problem, we consider an extension of the Klystron experiment from Section \ref{sec:klystron} by analysing a cascaded system.  
Specifically, we look at a system consisting of two Klystron. 
We first find an optimal spacing between the two cavity openings such that the mutual coupling is maximised. 
This separation distance was estimated by analysing the Fourier-domain response of the system (as shown in Table \ref{tab:klsytron_FD}).
Here, this occurs when they are placed 15 cm apart along the feed tube. 
As in the previous run, a background axial magnetic field of strength 1.5 $T$ was applied and a 40 kV, 1.5 A beam was injected into the feed tube.
Since the cavities have the same dimensions, their resonance peaks coincided, allowing us to downshift the EM system from $4.1$ GHz to baseband.

Once again, the RF-excitation at each port was set to be a modulated Gaussian pulse.
The quality factor $Q$ was measured at each peak by locating the resonance peak and finding the half power points.
The specific values of the quality factor are reported in Table \ref{tab:klystron_data}.
We note that as in the previous experiment, ET-FEMPIC yields very similar results as a traditional PIC solve, but at a significantly smaller timestep count.

\section{Summary \label{sec:conclusions}}

In this paper, we have proposed a technique to greatly improve the computational performance of EM-FEMPIC codes for a class of narrowband high frequency devices. Using a Quasi-Helmholtz setup, we show that our method exactly satisfies charge conservation and achieves similar fidelity to a traditional FEMPIC solve. As an aside, we have spent considerable fraction of the paper examining nuances of charge conservation from a different perspective. The upshot of this discussion is that the underlying rubric of quasi-Helmholtz decomposition is always robust and immune null spaces that would otherwise corrupt the system.  Additionally, our results demonstrate real world speedups equal to the ratio between the frequency shift and the bandwidth.
In particular, the Klystron example used close to a tenth the number of timesteps compared to a traditional EM-FEMPIC solve.
This is before any kind of additional speed up for parallel processing is applied. 

\section*{Acknowledgments}

This work was supported by the Department of Energy Computational Science Graduate Fellowship under grant DE-FG02-97ER25308 and financial support from NSF via CMMI-1725278. The authors would also like to thank the HPCC Facility,
Michigan State University, East Lansing, MI, USA.


\section{Appendices}
\subsection{Matrices involved in the Quasi-Helmholtz decomposition \label{sec:qhmats}}
As a prelude, we define the sets $\mathcal{N}$, $\mathcal{E}$, $\mathcal{F}$ and $\mathcal{T}$ as the set of nodes, edges, faces and tets respectively having $N_{n}$, $N_{f}$, $N_{e}$ and $N_{t}$ elements.
The various submatrices used in describing the Quasi-Helmholtz framework in Section \ref{sec:quasihelmholtz} are as follows:
\begin{align}
        [\star_\epsilon]_{i,j} &= \langle \vb{W}^{(1)}_i(\vb{r}),\varepsilon\cdot\vb{W}^{(1)}_j(\vb{r}) \rangle;  i,j \in \mathcal{E} \\
        [\star_{\mu^{-1}}]_{i,j} &= \langle \vb{W}^{(2)}_i(\vb{r}),\mu^{-1}\cdot\vb{W}^{(2)}_j(\vb{r})\rangle; i,j \in \mathcal{F}\\
        [\star_{\rho}]_{i,j} &= \langle W^{(3)}_i(\vb{r}),W^{(3)}_j(\vb{r})\rangle; i,j \in \mathcal{T}
\end{align}
where $\mathbf{W}^{(1)}_{i}$, $\mathbf{W}^{(2)}_{i}$ and $W^{(3)}_{i}$ are the Whitney edge, face and volume basis functions respectively. 
Further, we define the following matrices:
\begin{subequations}
\begin{align}
        [\vb{M}_g]_{i,j} &= \langle \vb{W}^{(1)}_i (\vb{r}) , \nabla W^{(0)}_j (\vb{r})\rangle; i \in \mathcal{E}, j \in \mathcal{N} \label{eq:mg}\\
        [\vb{M}_c]_{i,j} &= \langle  \vb{W}^{(2)}_i (\vb{r}), \curl \vb{W}^{(1)}_j (\vb{r})\rangle;  i \in \mathcal{F}, j \in \mathcal{E} \label{eq:mc}\\
        [\vb{M}_d]_{i,j} &= \langle W^{(3)}_i(\vb{r}) , \div \vb{W}^{(2)}_j (\vb{r}) \rangle;  i \in \mathcal{T}, j \in \mathcal{F} \label{eq:mc}\\
        [\grad ] &= \varepsilon[\star_{\epsilon}]^{-1}[\vb{M}_g] \label{eq:grad_mat}\\
        [\curl ] &= \mu^{-1}[\star_{\mu^{-1}}]^{-1}[\vb{M}_c]\label{eq:curl_mat}\\
        [\div ] & = [\star_{\rho}]^{-1}[\vb{M}_d] \label{eq:div_mat}
\end{align}
\end{subequations}
Likewise the submatrices involved in \eqref{eq:sol_disc} are as follows:
where the various submatrices involved in \eqref{eq:sol_disc} are defined as:
\begin{subequations}
\begin{align}
    [\vb{Z}]_{11} = &  [\vb{C}_c^b]^T[\vb{P}]_b^\Lambda [\vb{C}_c^b] \\
    [\vb{Z}]_{12} = & [\vb{C}_c^b]^T [\curl] [\star_\varepsilon]^{-1}[\vb{P}]_e^\Lambda [\star_\varepsilon] [\vb{C}_c^e] \\
    [\vb{Z}]_{13} = & [\vb{C}_c^b]^T [\curl] [\star_\varepsilon]^{-1}\Sigma [\vb{C}_z^e]\\
    [\vb{Z}]_{21}  = & [\vb{C}_c^e]^{T}[\vb{P}]_e^\Lambda[\star_\varepsilon][\vb{C}_c^e] \\
    [\vb{Z}]_{22} = & [\vb{C}_c^e]^{T} [\curl]^T [\star_{\mu^{-1}}][\vb{P}]_b^\Lambda [\vb{C}_c^b] \\
    [\vb{Z}]_{23} = & [\vb{C}_c^e]^{T} \Sigma  [\vb{C}_z^e]
\end{align}
\end{subequations}
where the $[C]_{b}$ matrices are mappings that identify unknowns that reside on the cotree. Constructing this mapping is trivial for simply connected structures, but is trickier for multiply connected geometries.
\subsection{Appendix B: Incompatibility of $W$ and $\delta$-testing \label{sec:Wn_innerprods}}
Suppose $\phi\left(t\right)=\tilde{\phi}\left(t\right)e^{j\omega t}$.
We can represent this as
\begin{equation}
\phi\left(t\right)=\sum_{i=n-1}^{n+1}\tilde{\phi}^{i}N_{i}\left(t\right)e^{j\omega t}
\end{equation}
where $N_{i}\left(t\right)$ reprasents the regular second order Newmark
basis functions. $\rho$ is defined as is normally done in Newmark:
\begin{equation}
\rho\left(t\right)=\sum_{i=n-1}^{n+1}\rho^{i}N_{i}\left(t\right)
\end{equation}
Now, $\partial_{t}\phi-\partial_{t}\rho$ becomes
\begin{equation}
\sum_{i=n-1}^{n+1}\left[\tilde{\phi}^{i}\partial_{t}\left(N_{i}\left(t\right)e^{j\omega t}\right)+\rho^{i}\partial_{t}N_{i}\left(t\right)\right]=0
\end{equation}
$W$-testing gives us
\begin{equation}
\begin{split}
&\left\langle W\left(t\right),\sum_{i=n-1}^{n+1}\left[\tilde{\phi}^{i}\partial_{t}\left(N_{i}\left(t\right)e^{j\omega t}\right)+\rho^{i}\partial_{t}N_{i}\left(t\right)\right]\right\rangle \\
&=\left\langle W\left(t\right),\sum_{i=n-1}^{n+1}\tilde{\phi}^{i}\left(N_{i}\left(t\right)\partial_{t}\left(e^{j\omega t}\right)+e^{j\omega t}\partial_{t}N_{i}\left(t\right)\right)\right\rangle\\ 
&+\left\langle W\left(t\right),\sum_{i=n-1}^{n+1}\rho^{i}\partial_{t}N_{i}\left(t\right)\right\rangle =0
\end{split}
\end{equation}
Thus, there are $6$ relevant terms in the expansion of $\left\langle W,\phi\left(t\right)\right\rangle $.
They are as follows:
\begin{equation}
\begin{split}
& \left\langle W\left(t\right),\partial_{t}\left(N_{k}e^{j\omega t}\right)\right\rangle = e^{j\omega t_{n}} \\
&\sum_{k=n-1}^{n+1}\left[\alpha_{k,0}+\alpha_{k,\Delta_{t}}e^{j\omega\Delta_{t}}+\alpha_{k,-\Delta_{t}}e^{-j\omega\Delta_{t}}\right]    
\end{split}
\end{equation}
where
\begin{subequations}
\begin{align}
\label{eq:coeffs}
\alpha_{n+1,0}&=\frac{4j+2\omega\Delta_{t}}{2\omega^{2}\Delta_{t}^{3}} \\
\alpha_{n+1,-\Delta_{t}}&=-\frac{2j+\omega\Delta_{t}}{2\omega^{2}\Delta_{t}^{3}} \\
\alpha_{n+1,\Delta_{t}}&=\frac{-2j-3\omega\Delta_{t}+2j\left(\omega\Delta_{t}\right)^{2}+2\left(\omega\Delta_{t}\right)^{3}}{2\omega^{2}\Delta_{t}^{3}} \\
\alpha_{n,0}&=\frac{-4j-2j\left(\omega\Delta_{t}\right)^{2}}{\omega^{2}\Delta_{t}^{3}} \\
\alpha_{n,-\Delta_{t}}&=\frac{2j-2\omega\Delta_{t}}{\omega^{2}\Delta_{t}^{3}} \\
\alpha_{n,\Delta_{t}}&=\frac{2j+2\omega\Delta_{t}}{\omega^{2}\Delta_{t}^{3}} \\
\alpha_{n-1,0}&=\frac{4j-2\omega\Delta_{t}}{2\left(\omega\Delta_{t}\right)^{3}} \\
\alpha_{n-1,-\Delta_{t}}&=\frac{-2j+3\omega\Delta_{t}+2j\left(\omega\Delta_{t}\right)^{2}-2\left(\omega\Delta_{t}\right)^{3}}{2\left(\omega\Delta_{t}\right)^{3}} \\
\alpha_{n-1,\Delta_{t}}&=\frac{-2j-\omega\Delta_{t}}{2\left(\omega\Delta_{t}\right)^{3}}
\end{align}
\end{subequations}
We can conclude by inspection from the coefficients defined in \eqref{eq:coeffs} that for general $\omega$, all nine coefficients can be nonzero.
As a result, the quantities $\beta_{k}=\left[\alpha_{k,0}+\alpha_{k,\Delta_{t}}e^{j\omega\Delta_{t}}+\alpha_{k,-\Delta_{t}}e^{-j\omega\Delta_{t}}\right]$
defined for $k=\left\{ n-1,n,n+1\right\}$ must also generally by nonzero. Then,
\begin{equation}
\begin{split}
\left\langle W\left(t\right),\partial_{t}\phi-\partial_{t}\rho\right\rangle &= \beta_{n+1}\phi^{n+1}+\beta_{n}\phi^{n}+\\
&\beta_{n-1}\phi^{n-1}e^{j\omega t_{n}}-\left(\frac{\rho^{n+1}-\rho^{n-1}}{2}\right)=0
\end{split}
\end{equation}
cannot be captured in a pointwise manner, since all three coefficients multiplying $\phi^{k}$ can generally by nonzero, while the coefficients multiplying $\rho^{k}$ \emph{only} exist when $k=\left\{n-1,n+1\right\}$.
\bibliographystyle{IEEEtran}
\bibliography{HOtime}

\end{document}